\documentclass[11pt]{article}
\usepackage{graphicx}
\usepackage{caption}
\usepackage{subfigure}
\usepackage{physics}
\usepackage{indentfirst}
\usepackage[section]{placeins}
\usepackage{float}
\usepackage{color}
\usepackage{cite}
\usepackage{multirow}
\usepackage{amssymb}
\begin{document}
	
\begin{center}
\large \textbf{Quasinormal modes and the correspondence with shadow in black holes with a deficit solid angle and quintessence-like matter}
\end{center}

\begin{center}
Chengye Yu $^{a}$\footnote{ E-mail: {chengyeyu1@hotmail.com}},
Deyou Chen $^{a}$\footnote{ E-mail: {deyouchen@hotmail.com}},
Chuanhong Gao $^{a}$\footnote{ E-mail: {chuanhonggao@hotmail.com}}
\end{center}

\begin{center}
$^{a}$ School of Science, Xihua University, Chengdu 610039, China
\end{center}

{\bf Abstract:} In this paper, we investigate the photon sphere, shadow radius and quasinormal modes of a $4$-dimensional black hole with a deficit solid angle and quintessence-like matter. We find that the radii of the photon sphere and shadow decrease with the decreases of the deficit solid angle and density of quintessence-like matter. The quasinormal modes are gotten by the sixth order WKB approximation method and shadow radius, respectively. The values of the real part and imaginary parts of the quasinormal modes increase with the decrease of the values of the deficit solid angle and density of quintessence-like matter when the multipole number is fixed. The quasinormal modes gotten by these two methods are in good agreement, especially when the multipole number is large. It shows the correspondence between the quasinormal modes in the eikonal limit and shadow.


\section{Introduction}

Quasinormal modes (QNMs) are asymptotic solutions of perturbation fields around compact objects under certain boundary conditions and characterize discontinuous complex frequencies. Their real parts determine oscillation frequencies and the imaginary parts describe the attenuation characteristics of the oscillation. Because these frequencies are determined by the space-time structure of the compact objects, the research on QNMs is helpful to understand the internal information of the compact objects. The discovery of gravitational wave signals of black holes is an evidence of the existence of the black holes. Recently, through the observation of the QNMs by the LIGO/Virgo laboratories\cite{BPA}, people found the gravitational wave signals, which shows the existence of the black holes. Another reason involves the AdS/CFT correspondence\cite{JMM,HH1,HH2,KSS,MM}. It shows that the QNMs of a $(D+1)$-dimensional asymptotically anti-de Sitter spacetime are poles of the retarded Green function in the $D$-dimensional dual conformal field theory. This correspondence has been successfully applied to the research of various properties of strongly coupled quark-gluon plasmas.

Through perturbations of various fields, the QNMs were obtained by various methods\cite{BBM,KDBF,ESSI,HPN1,HPN2,CCDN1,CCDN2,LLQ,BKK1,BKK2,BKK3,BKK4,BKK5,BKK6,BKK7,RVA}. The continued fraction method\cite{EW,EW1,EW2}, phase integral method\cite{NFPO,NASL,DVAA}, Hill-determinant method\cite{BMNP} and WKB approximation method are well-known methods and share different accuracy. The continued fraction method is employed and related to the three term recurrence relation. There is a possibility to calculate high overtones by using this method. The WKB approximation method was first proposed by Schutz and Will \cite{SW}, and then extended to the third-order \cite{IW}, sixth order and higher-order cases\cite{IW1,IW2}. The accuracy of the QNMs depends on the order of the WKB approximation. The properties of QNMs can be reflected by properties of null particles trapped at the unstable circular orbit. In\cite{CMBWZ},  Cardoso et al. pointed out that the real parts are determined by the product of the angular velocity of unstable orbit and multiple number, and the imaginary parts are determined by the overtone number and Lyapunov exponent. However, this derivation may not be fulfilled in the gravitational and other non-minimally coupled fields\cite{RAZS,RAAF}.

An important advance in astronomical observations is the derivation of the first image of the black hole in the center of galaxy M87 captured by the Event Horizon Telescope Collaboration\cite{EHT}. This image describes the presence of a crescent like bright area in the dark background and reveals the shapes of shadow and photon ring. This work provides an important observational proof for the existence of the black hole in our universe. As an observable quantity, shadow radii can be expressed by celestial coordinates. In the recent work \cite{JAAM1,JAAM2}, Jusufi further proved that real parts of QNMs in the eikonal limit can be given by shadow radii of black holes. This work reflects the correspondence between the shadows and QNMs in the eikonal limit. This correspondence was verified by the QNMs obtained by the third-order WKB approximation method\cite{JAAM3,JAAM4} and the sixth-order WKB approximation method\cite{GM}, respectively.

In this paper, we investigate the photon sphere, shadow and QNMs of a 4-dimensional black hole with a deficit solid angle and quintessence-like matter\cite{MBAV}. As one of the candidates of dark energy, quintessence is essentially different from the cosmological constant model. In the cosmological constant model, the state parameter $\omega _{0}$ is fixed at $- 1$. While in the quintessence model, the parameter is in a range of $- 1 \le \omega _{0}<0$. Therefore, the cosmological constant can be treated as a special quintessence. Based on the quintessence, some exact black hole solutions were gotten and the thermodynamic properties were discussed\cite{BL1,BL2,BL4,BL3,TD2,OPLA,OPLA2,BTZB}. In the study of the perturbations of the state parameter, Chakrabarty et al found that the perturbations with high value the parameter are unstable\cite{HAC}. In axisymmetric black holes, the quintessence-like matter changes the shape of the black hole shadow \cite{CXYQ}. To verify the validity of calculating QNMs by shadow radii, we get QNMs by the sixth order WKB approximation method and shadow radius of the black hole, respectively. The QNMs derived by these two methods are in good agreement when the multipole number is large, which shows the correspondence between the QNMs in the eikonal limit and shadow.

The rest is organized as follows. In the next section, we discuss the relation between the radius of the photon sphere and deficit solid angle, density of quintessence-like matter. In Section III, we get the shadow radius by the photon sphere, and calculate QNMs by the shadow radius and WKB approximation method, respectively. The last section is devoted to our conclusions.

\section{Photon sphere of a black hole with a deficit solid angle and quintessence-like matter}

\subsection{Review the black hole with a deficit solid angle and quintessence-like matter}

To provide a reasonable explanation for the accelerated expansion of our universe, dark energy and dark matter were proposed. As one of these models, the scalar field dark energy model includes the quintessence, phantom, etc. Using the quintessence, Barriola and Vilenkin got a solution of a black hole surrounded by quintessence-like matter. This solution is given by \cite{MBAV}

\begin{eqnarray}
ds^{2}=-f(r)dt^{2}+\frac{1}{f(r)}dr^{2}+r^2(d\theta ^2+sin^2\theta d\phi^2),
\label{eq2.1}
\end{eqnarray}

\noindent where

\begin{eqnarray}
f(r)=1-\frac{2M}{r}-\varepsilon ^{2}+\frac{\rho_{0}}{3\omega _{0}}r^{-3\omega _{0}-1},
\label{eq2.2}
\end{eqnarray}

\noindent $M$ is the black hole mass, $\varepsilon ^{2}$ denotes a dimensionless parameter of a deficit solid angle, $\rho_{0}$ is the density of quintessence-like matter, and $\omega _{0}$ is the state parameter and satisfies $-1<\omega _{0}<-1/3$. When $\varepsilon ^{2} = \rho_{0} =0$, the above metric is reduced to the solution of Schwarzschild black holes. The black hole horizon is determined by $f(r)=0$. To discuss the solution of Eq. (\ref{eq2.2}), we rewrite it as $f(r)=-\frac{r^{-3\omega _{0}-1}}{3\omega _{0}}f_{\omega _{0}}(r)$, where

\begin{eqnarray}
f_{\omega _{0}}(r)=3\omega _{0}(\varepsilon ^{2}-1)r^{3\omega _{0}+1}+6M\omega _{0}r^{3\omega _{0}}-\rho_{0}.
\label{eq2.2.1}
\end{eqnarray}

\noindent Therefore, the solution of $f(r) = 0$ is reduced to that of $f_{\omega _{0}}(r) = 0$. The solution depends on the values of parameters $\omega _{0}$, $\varepsilon ^{2}$ and $\rho_{0}$. There are two roots for $f_{\omega _{0}}(r)=0$ when $\rho _{0}< (1-\varepsilon ^{2}) \left[\frac{(1-\varepsilon ^{2})(3\left | \omega _{0} \right |-1)}{6M\left | \omega _{0} \right |}\right]^{3\left | \omega _{0} \right |-1}$, which correspond to the event horizon $(r_{+})$ and cosmological horizon $(r_{c})$, respectively.  When $\rho _{0}\geq (1-\varepsilon ^{2}) \left[\frac{(1-\varepsilon ^{2})(3\left | \omega _{0} \right |-1)}{6M\left | \omega _{0} \right |}\right]^{3\left | \omega _{0} \right |-1}$, there is only one real root which denotes the cosmological horizon. The relation between $f_{\omega _{0}}(r)$ and $r$ is plotted in Figure 1. It should be noted here that when $r=2.44939$ and $r=280.268$, there is $f_{\omega _{0}}(r)=0$ for $\omega _{0}= -\frac{1}{2}$ in the left picture. This implies that $f_{\omega _{0}}(r)$ and the horizontal axis have two intersections with the increase of $r$ in the left picture. There are two real roots for $f_{\omega _{0}}(r)=0$, which correspond to event and cosmological horizons, respectively. There is a real root for $f_{\omega _{0}}(r)=0$ in the right picture.  

\begin{figure}[h]
	\centering
	\begin{minipage}[t]{0.48\textwidth}
		\centering
		\includegraphics[scale=0.63]{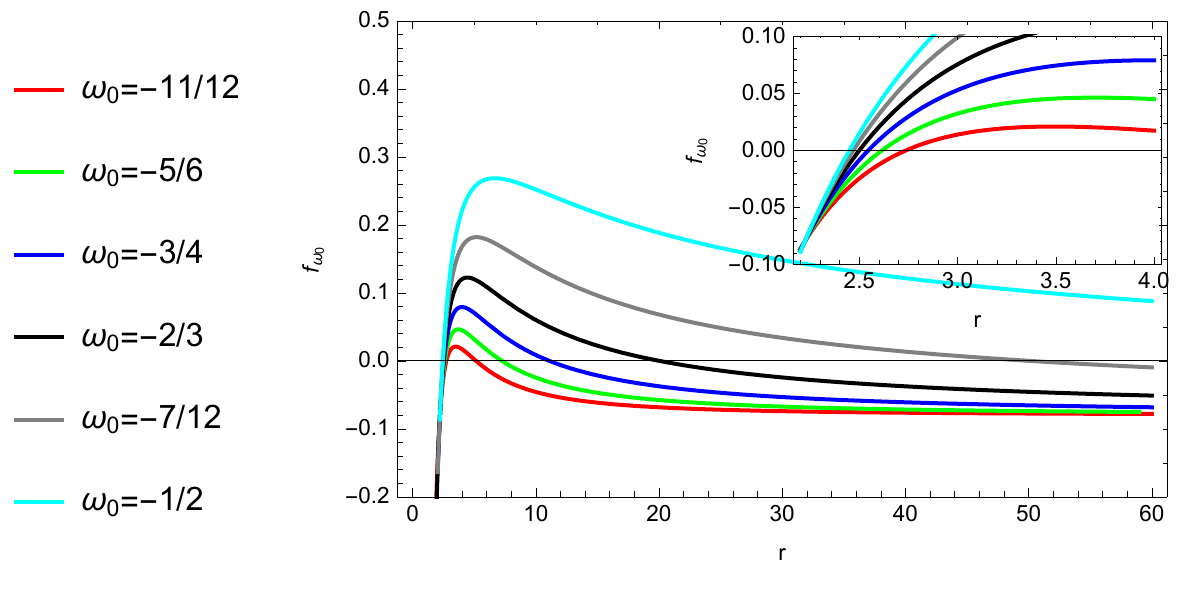}
	\end{minipage}
	\begin{minipage}[t]{0.48\textwidth}
		\centering
		\includegraphics[scale=0.46]{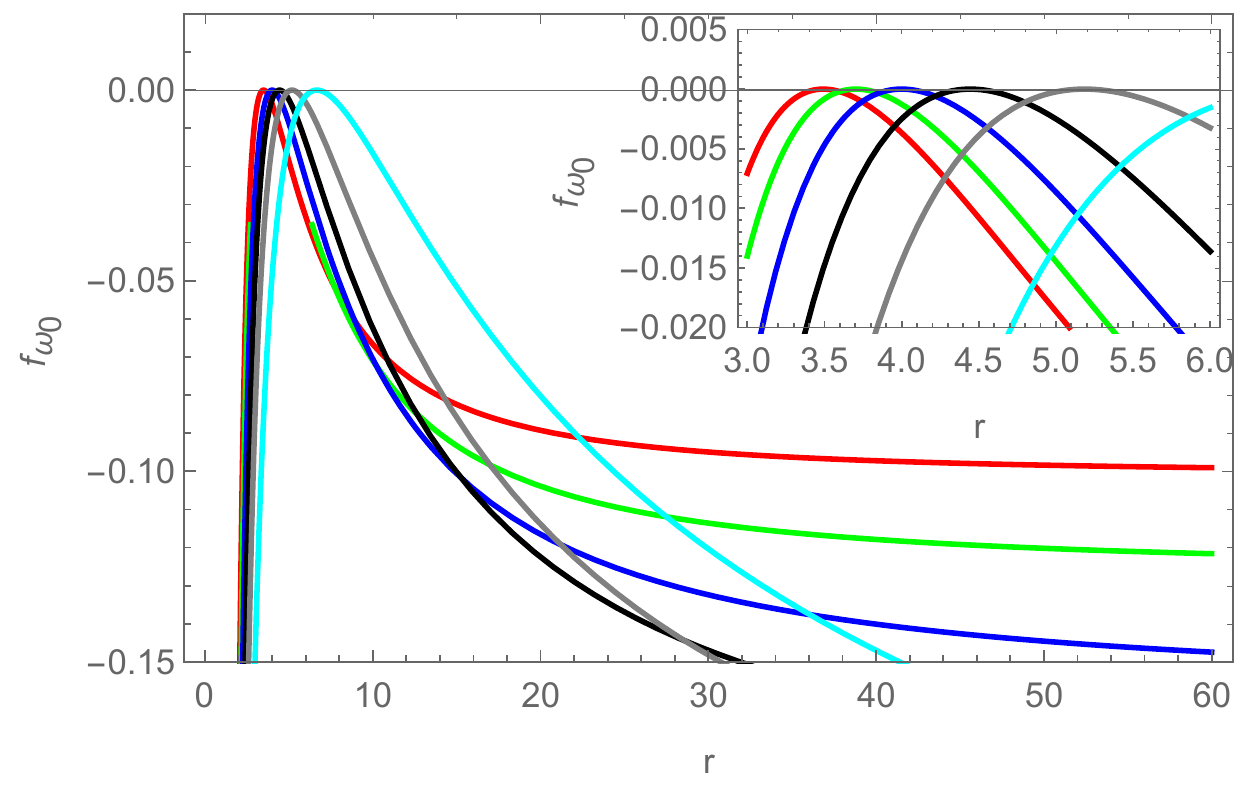}
	\end{minipage}
	\caption{The relation between $f_{\omega _{0}}(r)$ and $r$ is obtained, where $\rho _{0}< (1-\varepsilon ^{2})[\frac{(1-\varepsilon ^{2})(3\left | \omega _{0} \right |-1)}{6M\left | \omega _{0} \right |}]^{3\left | \omega _{0} \right |-1}$ $(\varepsilon ^{2}=0.1, \rho_{0}=0.08)$ is satisfied in the left picture and $\rho _{0}=(1-\varepsilon ^{2})[\frac{(1-\varepsilon ^{2})(3\left | \omega _{0} \right |-1)}{6M\left | \omega _{0} \right |}]^{3\left | \omega _{0} \right |-1}$ $(\varepsilon ^{2}=0.1)$ is satisfied in the right picture.}
	\label{fig:1.1}
\end{figure}

Although there are many values for $\omega _{0}$, we don't need to investigate all cases for these values. Without loss of generality, we focus our attention on the cases of $\omega _{0}=-\frac{2}{3}$ and $\omega _{0}=-\frac{1}{2}$. When $\omega _{0}=-\frac{2}{3}$, Eq. (\ref{eq2.2.1}) becomes $f_{\omega _{0}}(r)=-\frac{\rho _{0}}{r^{2}}(r-r_{+})(r-r_{c})$, where the cosmological and event horizons are located at
 
\begin{eqnarray}
r_{c}=\frac{1-\varepsilon ^{2}+\sqrt{(\varepsilon ^{2}-1)^{2}-4M\rho_{0}}}{\rho_{0}}, \quad r_{+}=\frac{1-\varepsilon ^{2}-\sqrt{(\varepsilon ^{2}-1)^{2}-4M\rho_{0}}}{\rho_{0}},
\label{eq2.3}
\end{eqnarray}

\noindent respectively. When the event and cosmological horizons coincide with each other, there is $\rho _{0}=(\varepsilon ^{2}-1)^{2}/4M$ and the black hole is an extremal one.

When $\omega _{0}=-\frac{1}{2}$, there are three roots for $f_{\omega _{0}}(r)=0$, which are

\begin{eqnarray}
r_{1}&=&\frac{3(1-\varepsilon ^{2})^{2}}{4\rho_{0}^{2}} +\frac{9(1-\varepsilon ^{2})^{4}-48(1-\varepsilon ^{2})M\rho_{0}^{2}}{4\rho_{0}^{2}\sqrt[3]{3\Xi }}+\frac{\sqrt[3]{3\Xi }}{4\rho_{0}^{2}},\nonumber\\
r_{2}&=&\frac{3(1-\varepsilon ^{2})^{2}}{4\rho_{0}^{2}}+\frac{(-3)^{\frac{2}{3}}[3(1-\varepsilon ^{2})^{4}-16(1-\varepsilon ^{2})M\rho_{0}^{2}]}{4\rho_{0}^{2}\sqrt[3]{\Xi }}-\frac{(1+\sqrt{3}i)\sqrt[3]{3\Xi }}{8\rho _{0}^{2}},\nonumber\\
r_{3}&=&\frac{3(1-\varepsilon ^{2})^{2}}{4\rho_{0}^{2}}-\frac{(1+\sqrt{3}i)[9(1-\varepsilon ^{2})^{4}-48(1-\varepsilon ^{2})M\rho_{0}^{2}]}{8\rho_{0}^{2}\sqrt[3]{\Xi }}-\frac{(1-\sqrt{3}i)\sqrt[3]{3\Xi }}{8\rho _{0}^{2}},
\label{eq2.4}
\end{eqnarray}

\noindent where

\begin{eqnarray}
\Xi =9(1-\varepsilon ^{2})^{6}-72(1-\varepsilon ^{2})^{3}M\rho_{0}^{2}+96M^2\rho_{0}^{4}
+16\sqrt{36M^4\rho_{0}^{8}-6(1-\varepsilon ^{2})^{3}M^3\rho_{0}^{6}}.
\label{eq2.5}
\end{eqnarray}

\noindent From the figure, it is clearly that there are two real roots when $\omega _{0}=-\frac{1}{2}$ \cite{BTZB,OPLA,OPLA2}. Although the imaginary number $i$ appears in Eq. (\ref{eq2.4}), because the value of Eq. (\ref{eq2.5}) is a complex number, substituting this value into Eq. (\ref{eq2.4}) produces two real numbers, which correspond to an event horizon ($r_2$) and a cosmological horizon ($r_1$), respectively. The third root is not a real number and is neglected.

To display the region of the values ($\varepsilon ^{2},\rho_0$) where the spacetime has two horizons, we plot Figure 2. In the figure, the value of $\rho_0$ decrease with the increase of $\varepsilon ^{2}$, and the region where two horizons exist is larger than that corresponding to one horizon.

\begin{figure}[h]
	\centering
	\begin{minipage}[t]{0.48\textwidth}
		\centering
		\includegraphics[scale=0.6]{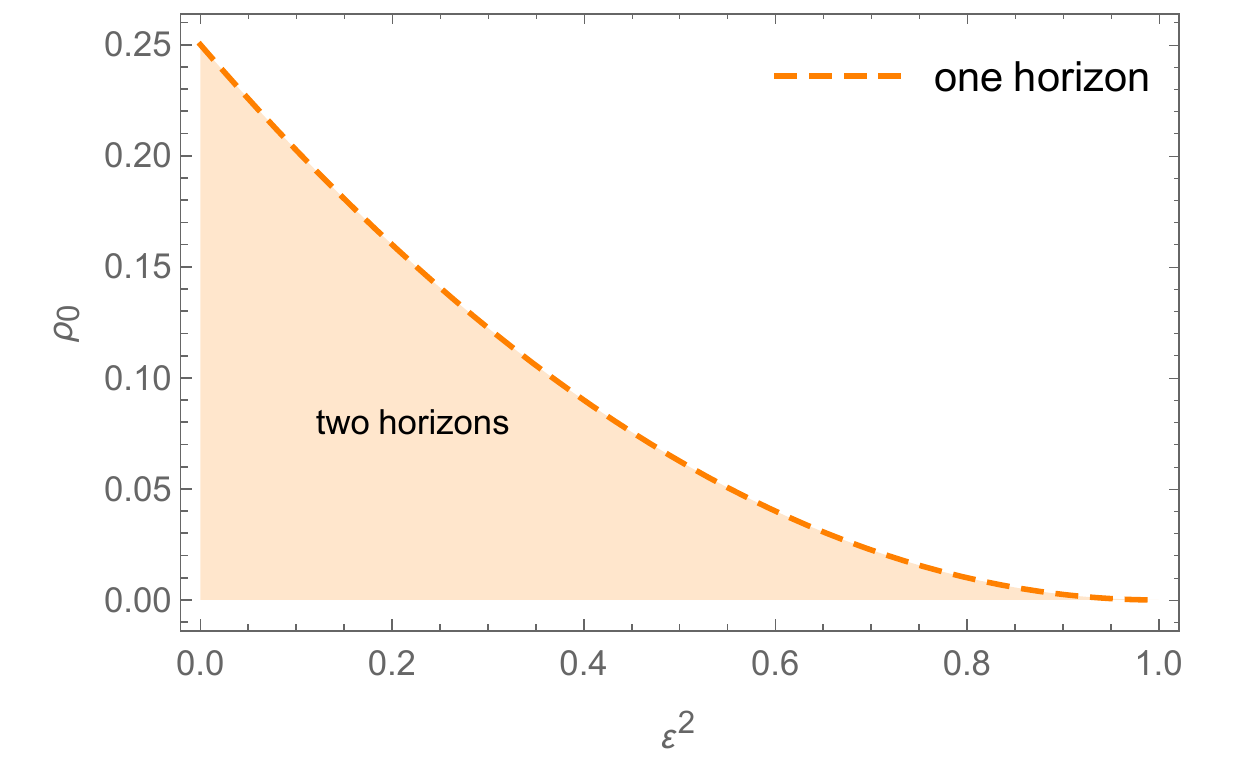}
	\end{minipage}
	\begin{minipage}[t]{0.48\textwidth}
		\centering
		\includegraphics[scale=0.6]{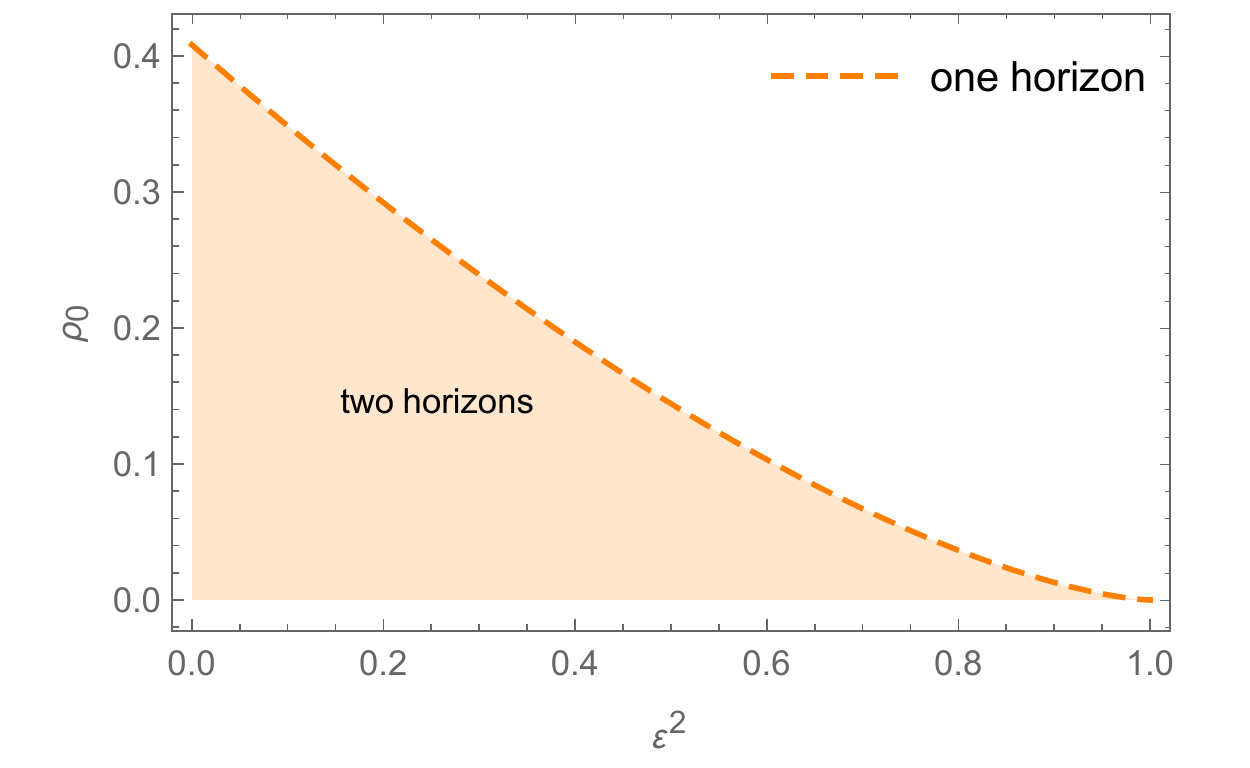}
	\end{minipage}
	\caption{The behavior of $\varepsilon^2$ as a function $\rho _{0}$ for $\omega_{0}=-\frac{2}{3}$ is plotted in the left picture, and for $\omega_{0}=-\frac{1}{2}$ is plotted in right picture. When $\rho _{0}= (1-\varepsilon ^{2})[\frac{(1-\varepsilon ^{2})(3\left | \omega _{0} \right |-1)}{6M\left | \omega _{0} \right |}]^{3\left | \omega _{0} \right |-1}$, there is only a horizon. }
	\label{fig:1.2}
\end{figure}

\subsection{Photon sphere}

We first consider a photon orbiting around the black hole (\ref{eq2.1}), where the photon moves along a null geodesic. At the equatorial orbit, $\theta = \frac{\pi}{2}$. Its Lagrangian is

\begin{eqnarray}
\mathcal{L} &=& \frac{1}{2}\left[-f(r)\dot{t}^2 + \frac{1}{f(r)}\dot{r}^2 + r^2\dot{\phi}^2\right].
\label{eq2.6}
\end{eqnarray}

\noindent In the above equation, the dot over a symbol is the differentiation with respect to an affine parameter. Using the definition of momenta and Lagrangian (\ref{eq2.6}), we get generalized momenta

\begin{eqnarray}
p_t &=& -f(r)\dot{t}=-E,\nonumber\\
p_r &=& \frac{\dot{r}}{f(r)},\nonumber\\
p_{\phi} &=& r^2\dot{\phi}=L.
\label{eq2.7}
\end{eqnarray}

\noindent where $E$ and $L$ denote the energy and orbital angular momentum of the photon, respectively. From Eqs. (\ref{eq2.6}), (\ref{eq2.7}) and the Hamiltonian for this system

\begin{eqnarray}
\mathcal{H} &=& 2\left(p_{\mu}\dot{x}^{\mu}- \mathcal{L} \right)= -E\dot{t} +\frac{\dot{r}^2}{f(r)} +L\dot{\phi}=0,
\label{eq2.8}
\end{eqnarray}

\noindent the equation of radial motion is gotten as

\begin{eqnarray}
\dot{r}^2 +\tilde{V}(r)=0,
\label{eq2.9}
\end{eqnarray}

\noindent where $\tilde{V}(r)=- E^2+ \frac{L^2}{r^2}f(r)$ is an effective potential of the photon and determines the position of the unstable orbit around the black hole. This orbit satisfies the following conditions

\begin{eqnarray}
\tilde{V}(r)=0, \quad\quad \frac{\partial \tilde{V}(r)}{\partial r} = 0, \quad\quad \frac{\partial^2 \tilde{V}(r)}{\partial r^2} <0.
\label{eq2.10}
\end{eqnarray}

\noindent Solving the middle equation of the above equations yields four roots. We use the first and third equations to exclude the other meaningless roots and get the radius of photon sphere. When $\omega _{0}=-\frac{2}{3}$, we use Eq. (\ref{eq2.10}) and get

\begin{eqnarray}
r_{ps}^{+}=\frac{2(1-\varepsilon ^{2})-\sqrt{4(\varepsilon ^{2}-1)^{2}-12M\rho_{0}}}{\rho_{0}}.
\label{eq2.11}
\end{eqnarray}

\noindent When $\omega _{0}=-\frac{1}{2}$, the photon sphere is located at

\begin{eqnarray}
r_{ps}^{+}=\frac{4(1-\varepsilon ^{2})^{2}}{3\rho_{0}^{2}}-\frac{-16(1-\varepsilon ^{2})^{4}+72(1-\varepsilon ^{2})M\rho_{0}^{2}}{3\rho_{0}^{2}\sqrt[3]{2\varpi }}+\frac{\sqrt[3]{2\varpi }}{3\rho_{0}^{2}},
\label{eq2.12}
\end{eqnarray}

\noindent where

\begin{eqnarray}
\varpi &=&32(1-\varepsilon ^{2})^{6}-216(1-\varepsilon ^{2})^{3}M\rho_{0}^{2}+243M^{2}\rho_{0}^{4}\nonumber\\
&+&27\sqrt{-16(1-\varepsilon ^{2})^{3}M^{3}\rho_{0}^{6}+81M^{4}\rho_{0}^{8}}.
\label{eq2.13}
\end{eqnarray}

\begin{figure}[h]
	\centering
	\begin{minipage}[t]{0.48\textwidth}
		\centering
		\includegraphics[scale=0.32]{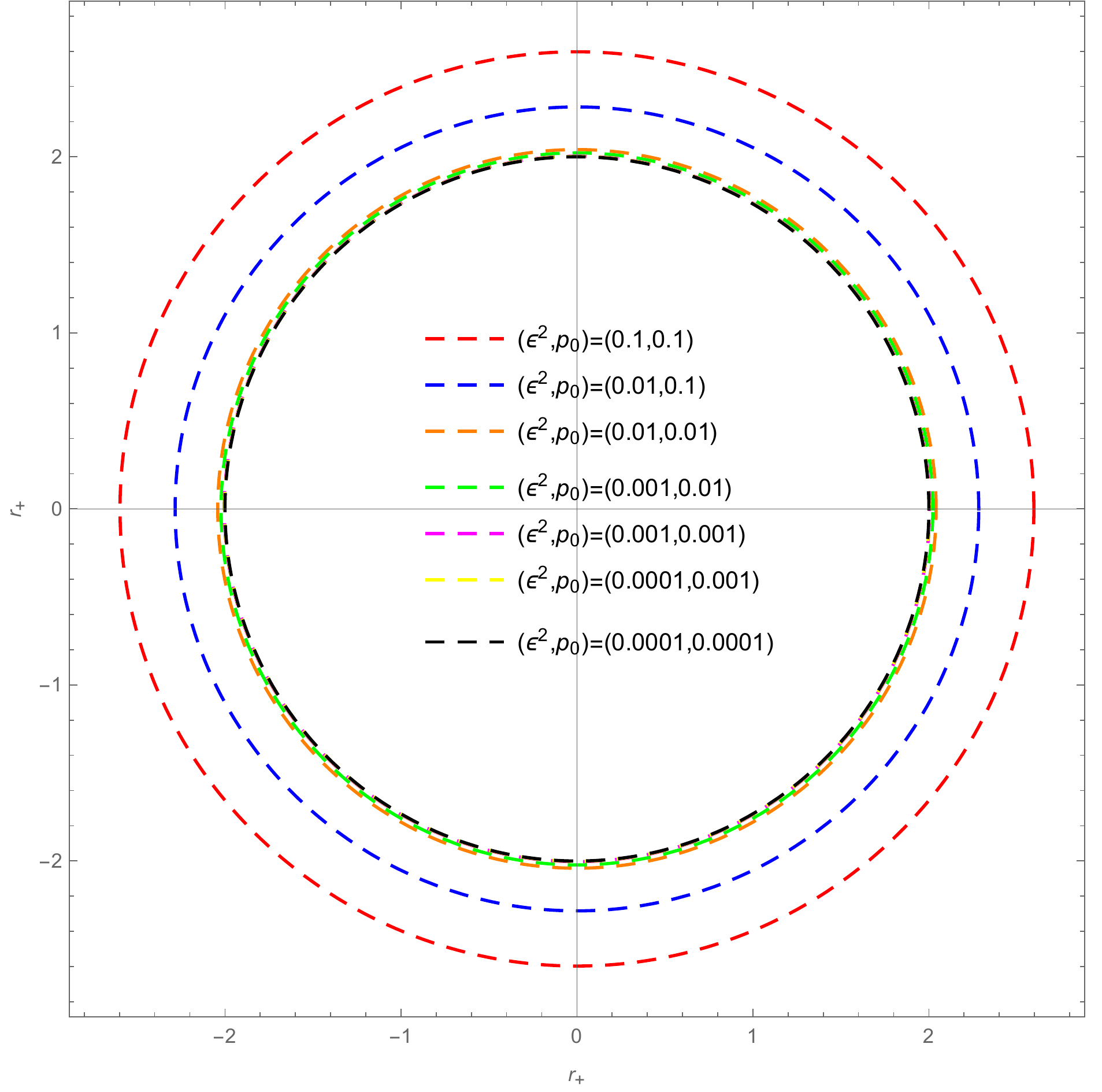}
	\end{minipage}
	\begin{minipage}[t]{0.48\textwidth}
		\centering
		\includegraphics[scale=0.335]{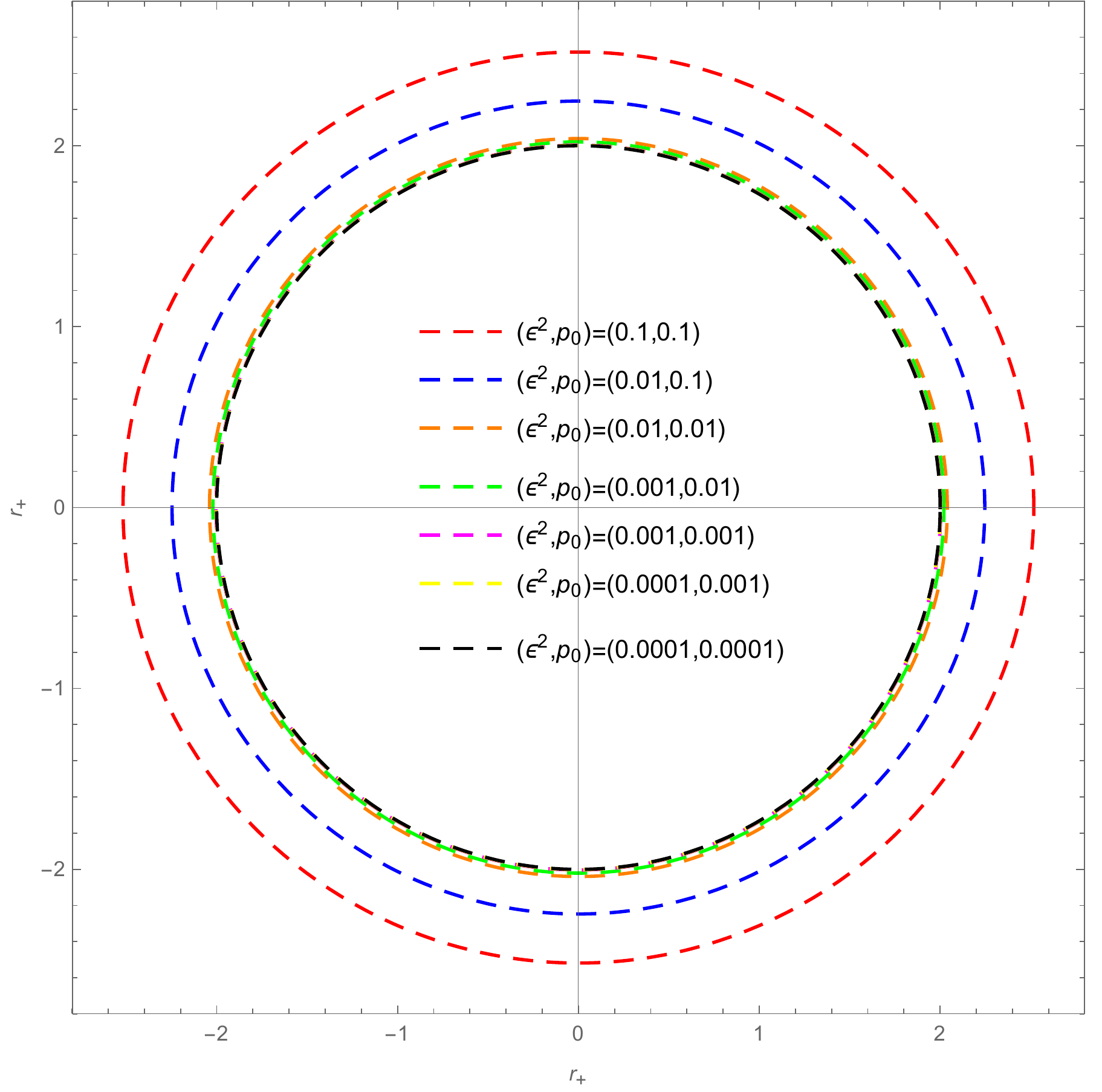}
	\end{minipage}
	\caption{The influence of ($\varepsilon ^{2}$,$\rho_{0}$) on the event horizon, where $M=1$ and $L=1$. The left picture is plotted at $\omega _{0}=-\frac{2}{3}$, and the right picture is plotted at $\omega _{0}=-\frac{1}{2}$.}
	\label{fig:2.3}
\end{figure}

\begin{figure}[h]
	\centering
	\begin{minipage}[t]{0.48\textwidth}
		\centering
		\includegraphics[scale=0.3]{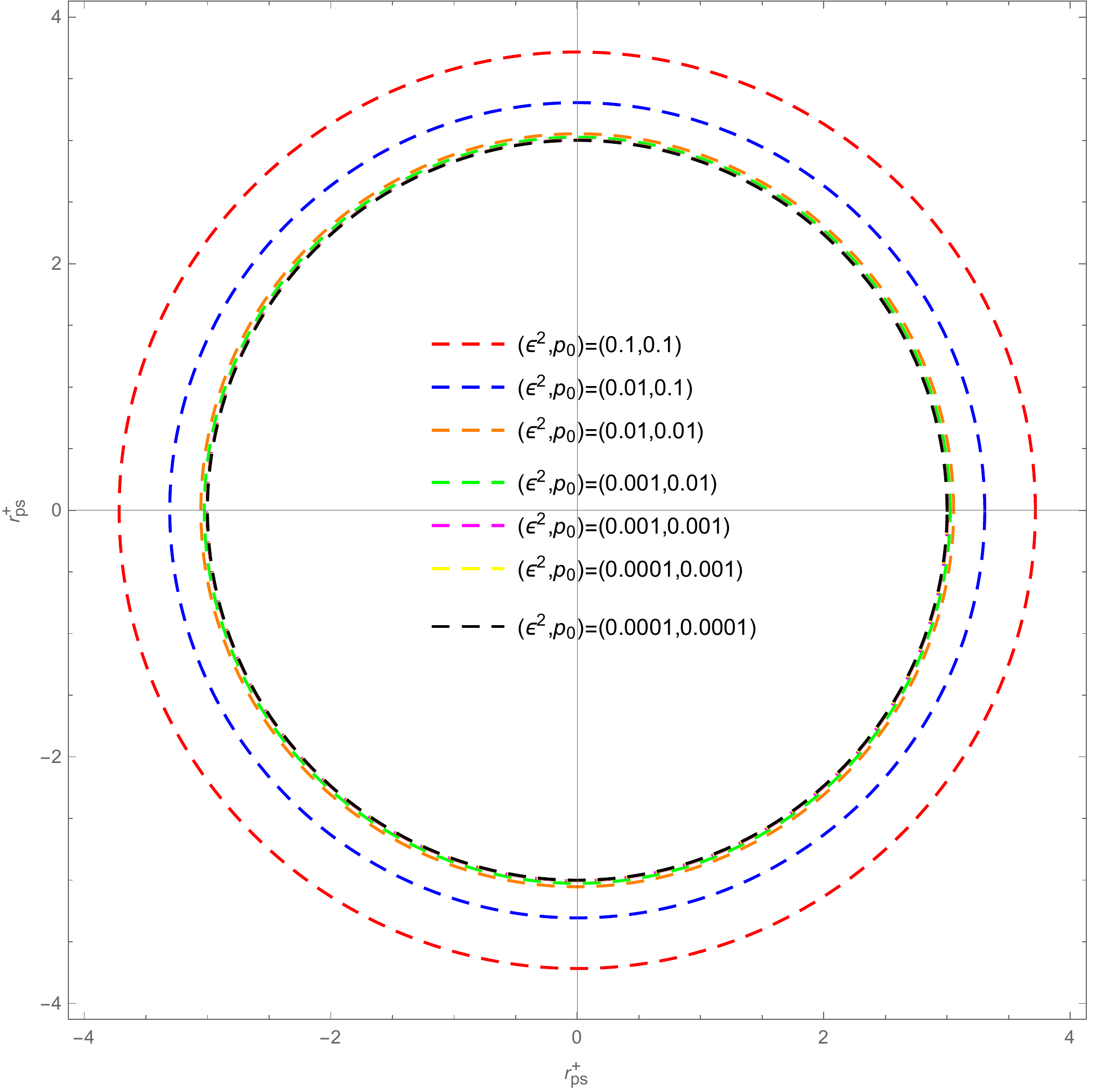}
	\end{minipage}
	\begin{minipage}[t]{0.48\textwidth}
		\centering
		\includegraphics[scale=0.3]{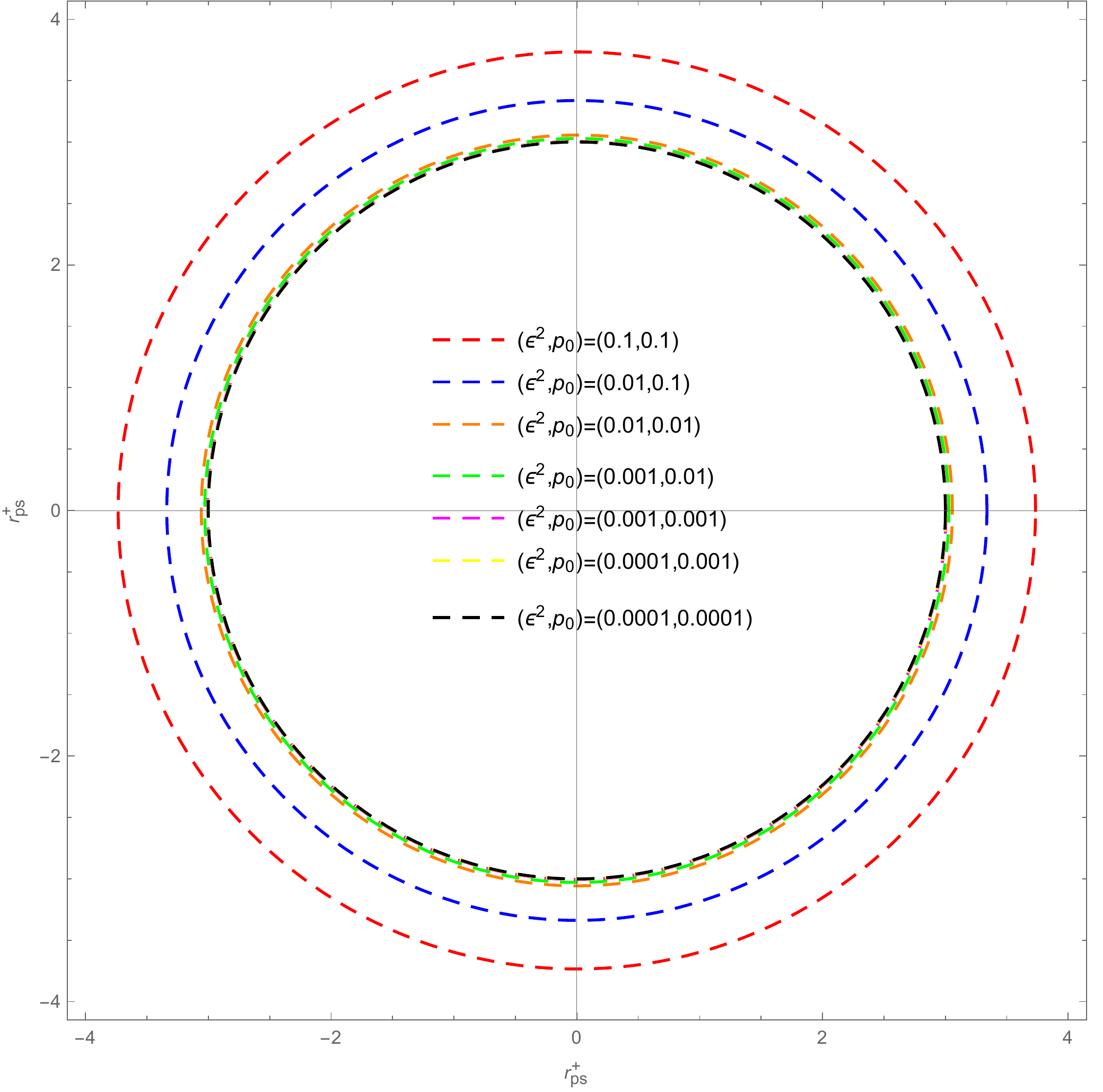}
	\end{minipage}
	\caption{The influence of ($\varepsilon ^{2}$,$\rho_{0}$) on the radius of the photon sphere, where $M=1$ and $L=1$. The left picture is plotted at $\omega _{0}=-\frac{2}{3}$, and the right picture is plotted at $\omega _{0}=-\frac{1}{2}$.}
	\label{fig:2.4}
\end{figure}

\noindent To be more intuitive, we plot the radii of the event horizon and photon sphere in Figure \ref{fig:2.3} and Figure \ref{fig:2.4}. From Figure \ref{fig:2.3}, it is found that when $\rho_{0}$ and $\varepsilon ^{2}$ decrease, the event horizon shrinks and tends to a limit value 2. From Figure \ref{fig:2.4}, we find that when $\rho_{0}$ and $\varepsilon ^{2}$ decrease, the photon sphere radius decreases and tends to a limit value 3.

\section{The shadow and QNMs }

The WKB approximation method is an effective way to derive QNMs. It was first proposed in \cite{SW}, and extended to the third order WKB approximation in \cite{IW} and higher order WKB approximation in \cite{IW1,IW2}. In this section, we use a scalar field perturbation in the spherically symmetrical black hole with quintessence-like matter and a deficit solid angle to derive the QNMs by the sixth order WKB approximation method.

The massless scalar field in curved spacetime obeys

\begin{eqnarray}
\frac{1}{\sqrt{-g}}\partial_{\mu}\left(\sqrt{-g}g^{\mu\nu}\partial_{\nu}\Psi\right)=0.
\label{eq3.1}
\end{eqnarray}

\noindent We use the following ansatz

\begin{eqnarray}
\Psi =\sum_{l,m}{\frac{e^{-i\omega t}}{r}}\Phi (r)Y_{lm}(\theta ,\varphi ),
\label{eq3.2}
\end{eqnarray}

\noindent where $e^{-i\omega t}$ denotes the time evolution of the scalar field,  $\omega$ is the perturbation frequency and $Y_{lm}(\theta, \phi)$ is a spherical harmonics. Inserting Eq. (\ref{eq3.2}) and the inverse metric components of Eq. (\ref{eq2.1}) into Eq. (\ref{eq3.1}) and performing a "tortoise" coordinate transformation $dr_{\star}=\frac{dr}{f(r)}$ yield

\begin{eqnarray}
\frac{d^{2}\tilde{\Phi }}{dr_{*}^{2}}+(\omega ^{2}-V)\tilde{\Phi }(r_{*})=0,
\label{eq3.3}
\end{eqnarray}

\noindent where $\omega $ is the complex quasinormal frequencies, and $V$ is the effective potential of the perturbation field. For an asymptotically flat spacetime, Eq. (\ref{eq3.3}) is solved by combining with the condition,  

\begin{eqnarray}
\frac{\omega ^{2}-V}{\sqrt{2(\omega ^{2}-V)^{(2)}}}=i(n+\frac{1}{2}),
\label{eq3.3.1}
\end{eqnarray}

\noindent and $(\omega ^{2}-V)^{(2)}\equiv \partial^2 (\omega ^{2}-V)/\partial r_{*}^2$. And then the effective potential for the scalar perturbation is obtained 

\begin{eqnarray}
V(r)=f(r)\left[\frac{f'(r)}{r}+\frac{L(L+1)}{r^{2}}\right].
\label{eq3.4}
\end{eqnarray} 

\begin{figure}[h]
	\centering
	\begin{minipage}[t]{0.48\textwidth}
		\centering
		\includegraphics[scale=0.55]{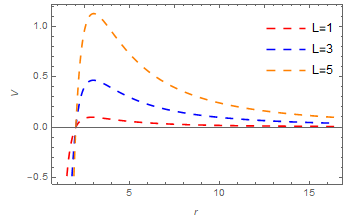}
	\end{minipage}
	\begin{minipage}[t]{0.48\textwidth}
		\centering
		\includegraphics[scale=0.55]{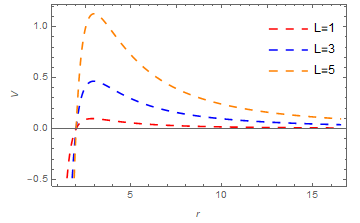}
	\end{minipage}
	\caption{The behavior of $V(r)$ for different values of $r$ and $L$, where $\rho_{0}=0.001$ and $\varepsilon  ^{2}=0.001$. In the left picture, $\omega _{0}=-\frac{2}{3}$. In the right picture,  $\omega _{0}=-\frac{1}{2}$.}
	\label{fig:3.1}
\end{figure}

\noindent In the above equation, $\prime$ denotes the differential of $r$. The behavior of $V(r)$ for different values of $r$ and $L$ is described in Figure \ref{fig:3.1}. In this paper, it is ordered that $M=1$. Using boundary conditions $r_{\star} \to \pm \infty$ which map the event horizon and infinity, we obtain the solution of Eq. (\ref{eq3.3}). It takes form $\Phi(r)\sim exp[-i\omega(t\mp)r_{\star}]$ with oscillation modes under the large-$l$ limit \cite{RVA,CMBWZ}

\begin{eqnarray}
\omega =\omega_{R}-i\omega_{I}=\Omega l-i\lambda (n+\frac{1}{2}),
\label{eq3.5}
\end{eqnarray}

\noindent where

\begin{eqnarray}
\Omega =\left.\sqrt{\frac{f^{\prime}(r)}{2r}}\right|_{r=r^{+}_{ps}}, \quad \quad \lambda=\left.\sqrt{\frac{V^{\prime\prime}(r)}{2\dot{t}^2}}\right|_{r=r^{+}_{ps}},
\label{eq3.6}
\end{eqnarray}

\noindent $\Omega $ is the angular velocity of the photon sphere at the unstable orbit, $l$ is the angular momentum, $n$ is the overtone number and $\lambda $ is Lyapunov exponent. It is not difficult to see from Eqs. (\ref{eq3.5}) and (\ref{eq3.6}) that QNMs can be obtained by properties of photon spheres. According to Eqs. (\ref{eq2.2}), (\ref{eq3.4}) and (\ref{eq3.5}), we use the sixth order WKB approximation method, and derive the values of the QNMs in Table 1-Table 12.  Here we only list the fundamental modes $(n=0)$, since they have the least damping among the detected ringing signals and dominate the waveform of gravitational waves.

Radii of shadows and photon spheres are closely related to parameters of black holes. In \cite{JAAM1,JAAM2}, Jusufi proved that the shadow radii can be expressed by the radii of photon spheres. For a spherically symmetrical black hole, the shadow radius is given by

\begin{eqnarray}
R_{sh}=\eval{\frac{r}{\sqrt{f(r)}}}_{r_{ps}^{+}}.
\label{eq3.7}
\end{eqnarray}

\begin{figure}[h]
	\centering
	\begin{minipage}[t]{0.48\textwidth}
		\centering
		\includegraphics[scale=0.34]{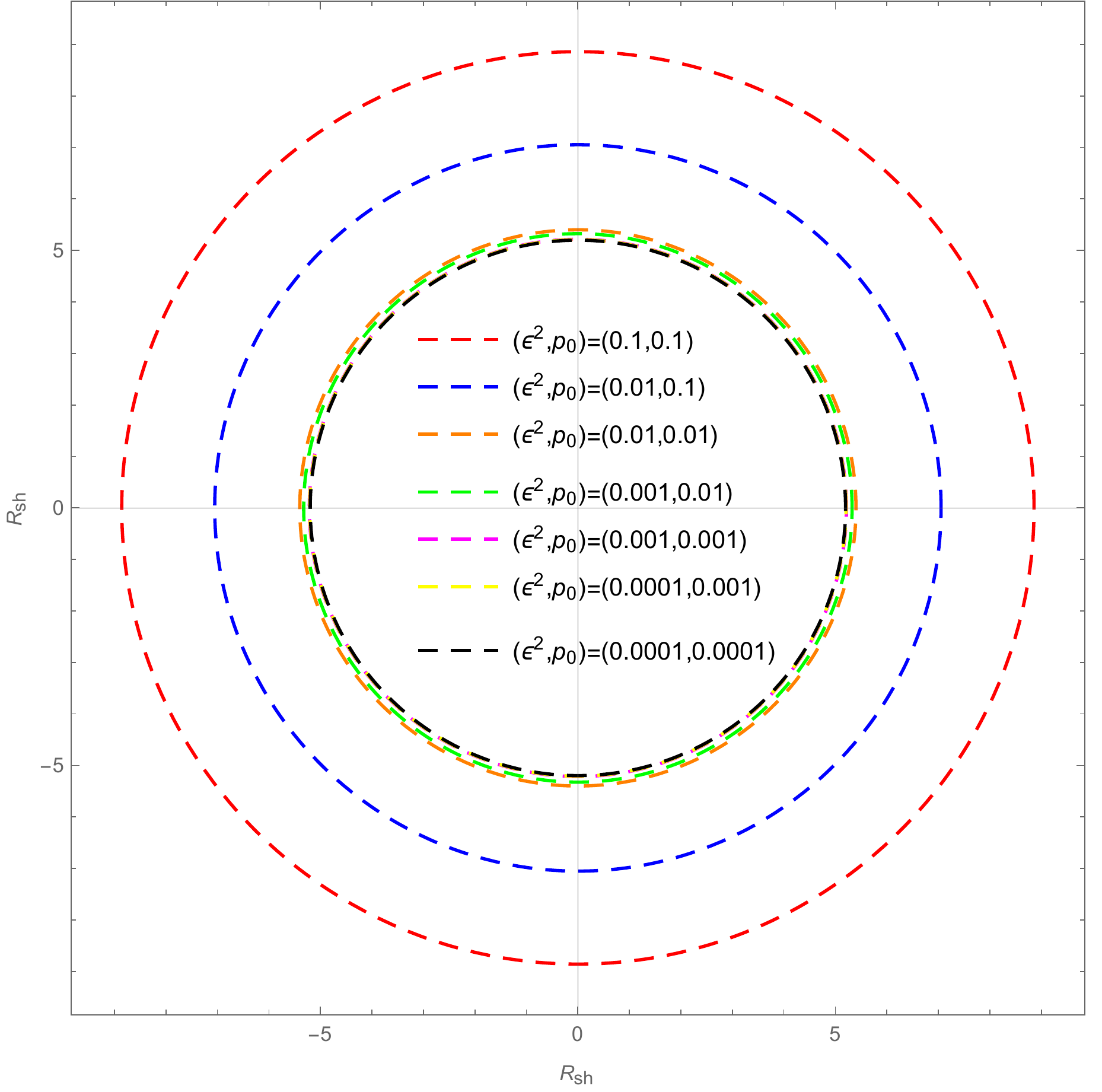}
	\end{minipage}
	\begin{minipage}[t]{0.48\textwidth}
		\centering
		\includegraphics[scale=0.34]{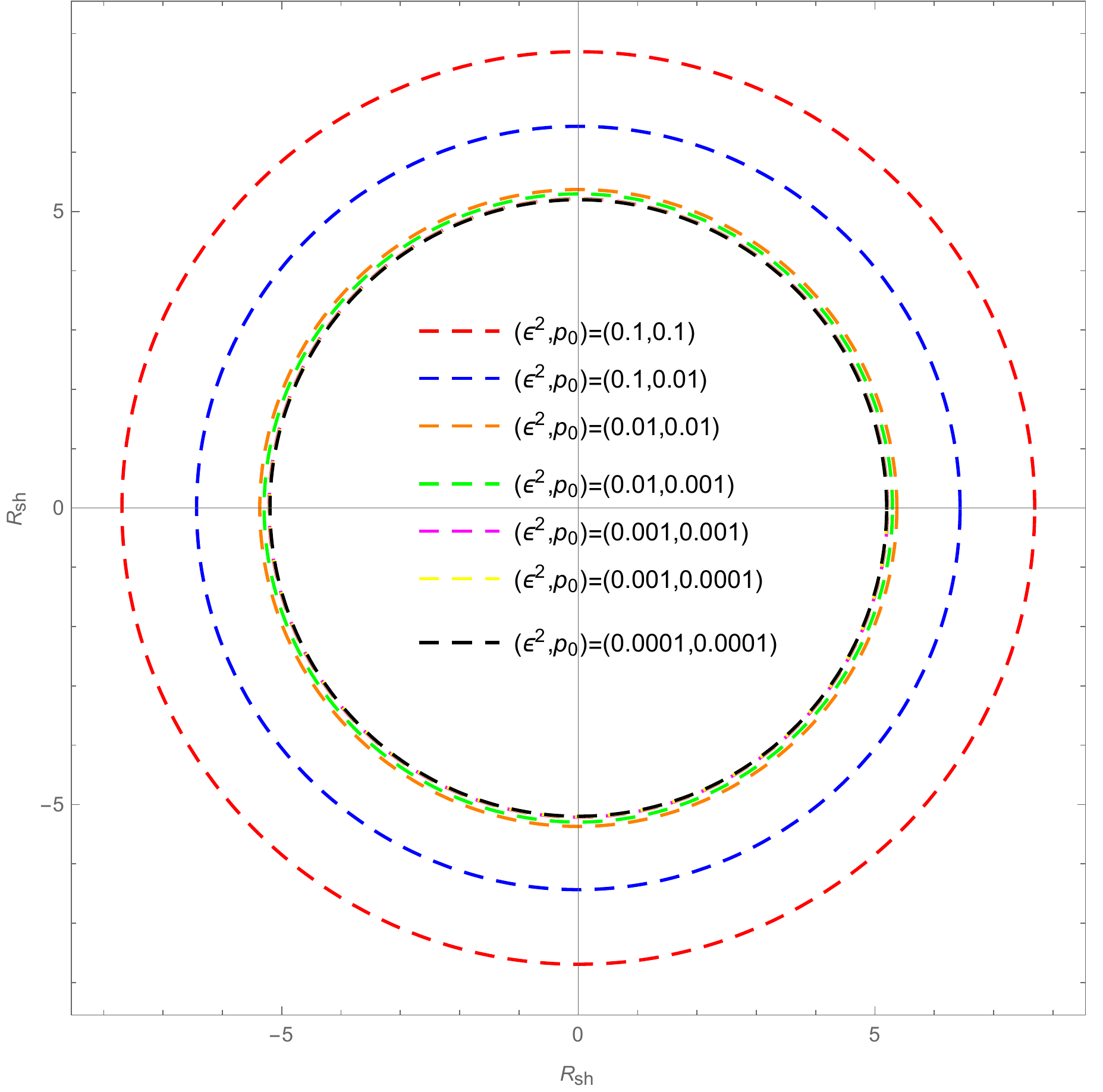}
	\end{minipage}
	\caption{The shadow radius is plotted in the left picture at $\omega _{0}=-\frac{2}{3}$ and in the right picture at $\omega _{0}=-\frac{1}{2}$, where $L=1$.}
	\label{fig:3.2}
\end{figure}

\noindent Considering the complexity of expression of the shadow radius, we describe its radius directly by two graphs in Figure \ref{fig:3.2}. From the figure, it is found that the shadow radius decreases with the decreases of variables $\rho_{0}$ and $\varepsilon ^{2}$ and approaches to a limit value at $5$.

\begin{figure}[h]
	\centering
	\begin{minipage}[t]{0.48\textwidth}
		\centering
		\includegraphics[height=4.5cm,width=7.3cm]{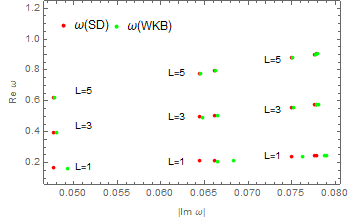}
	\end{minipage}
	\begin{minipage}[t]{0.48\textwidth}
		\centering
		\includegraphics[height=4.5cm,width=7.3cm]{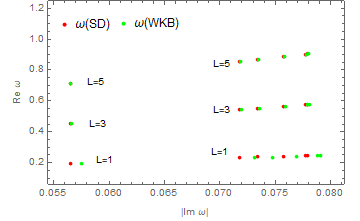}
	\end{minipage}
	\caption{The influence of $\rho_{0}$ and $\varepsilon ^{2}$ on the real and imaginary parts of the QNMs where $L=1,3,5$. The left picture is plotted at $\omega _{0}=-\frac{2}{3}$, and the right picture is plotted at $\omega _{0}=-\frac{1}{2}$.}
	\label{fig:3.3}
\end{figure}

Taking into account the relationship between the shadow and photon sphere, Jusufi further derived the QNMs in the eikonal limit by the shadow radius, which is

\begin{eqnarray}
\omega =\frac{2L+D-3}{2R_{sh}}-i\lambda (n+\frac{1}{2}).
\label{eq3.8}
\end{eqnarray}

\noindent where $D$ is a spacetime dimension, $L$ is the multiple number, and $\lambda$ is given by the Lyapunov exponent. Using Eq. (\ref{eq3.7}), we get the fundamental modes in Table 1-Table 12. In the tables, the relative deviations refer to the deviations of the values of the QNMs calculated by the shadow radius from those calculated by the WKB approximation method. In Figure \ref{fig:3.3}, we plot the relation between the real and imaginary parts of the QNMs with different $\rho_{0}$ and $\varepsilon ^{2}$. From the figure, we find that when $L=3$ or greater, the real and imaginary parts of QNMs calculated by the shadow radius and sixth order WKB are consistent. When $L=1$, the real and imaginary parts of QNMs calculated by these two approaches have certain deviations. The values of relative deviation are listed in Table 1-12.

From the above figures and Table 1-12, we find the following phenomena.

1. When the values of the density of quintessence-like matter and multiple number are fixed, the values of the real and imaginary parts of the QNMs increase with the decrease of the value of the deficit solid angle. The relative deviations decrease with the decease of the deficit solid angle.

2. When the values of the deficit solid angle and multiple number, the real and imaginary parts of the QNMs increase with the decrease of the density of quintessence-like matter. The relative deviations decrease with the decrease of the density of quintessence-like matter.

3. For the certain values of the deficit solid angle and density of quintessence-like matter, the values of the QNMs obtained by the two methods are more and more consistent with the increase of the value of the angular quantum number, which shows the correspondence between the QNMs in the eikonal limit and shadow. However, for a small multiple number, the relative deviations are also small.

4. For the fixed values of the deficit solid angle, density of quintessence-like matter and multiple number, the real and imaginary parts of the QNMs calculated at $\omega _{0}= -2/3$ is smaller than those calculated at $\omega _{0}= -1/2$. This implies that the QNMs calculated at $\omega _{0}= -2/3$ are more significant because they are more stable. For the cases $\omega _{0}\rightarrow -\frac{1}{3}$ and $\omega _{0}\rightarrow -1$, we infer that the real and imaginary parts of the former QNMs are larger than the latter. Therefore, the QNM frequencies gotten in the neighborhood of $\omega _{0}= -1$ is more stable than those gotten at $\omega _{0}= -2/3$ and $\omega _{0}= -1/2$.

\begin{table}[htbp]
	\centering
	\caption{The QNMs calculated by the shadow radius and sixth order WKB approximation method and the relative deviations, where $\omega _{0}=-\frac{2}{3}$, $\rho _{0}=0.1$, $\varepsilon ^{2}=0.1$.}
	\begin{tabular}{|c|c|c|c|c|}
		\hline
		&\multicolumn{2}{|c|}{QNMs }&\multicolumn{2}{c|}{Relative deviations}\\
		\hline
		$L$&$\omega (SD)$&$\omega (WKB)$&$\left | Re \omega  \right |$&$\left | Im \omega  \right |$\\
		\hline
		1&0.1693385-0.0477010i&0.1639632-0.0492834i&3.278\%&3.211\%\\
		\hline
		3&0.3951232-0.0477010i&0.3927900-0.0479800i&0.594\%&0.581\%\\
		\hline
		5&0.6209079-0.0477010i&0.6194204-0.0478134i&0.240\%&0.235\%\\
		\hline
		7&0.8466926-0.0477010i&0.8456013-0.0477613i&0.129\%&0.126\%\\
		\hline
		10&1.1853697-0.0477010i&1.1845900-0.0477317i&0.066\%&0.064\%\\
		\hline
		15&1.7498314-0.0477010i&1.7493032-0.0477151i&0.030\%&0.030\%\\
		\hline
		20&2.3142931-0.0477010i&2.3138937-0.0477090i&0.017\%&0.017\%\\
		\hline
		50&5.7010636-0.0477010i&5.7009015-0.0477023i&0.003\%&0.003\%\\
		\hline
		
	\end{tabular}
\end{table}

\begin{table}[htbp]
	\centering
	\caption{The QNMs calculated by the shadow radius and sixth order WKB approximation method and the relative deviations, where $\omega _{0}=-\frac{2}{3}$, $\rho _{0}=0.1$, $\varepsilon ^{2}=0.01$.}
	\begin{tabular}{|c|c|c|c|c|}
		\hline
		&\multicolumn{2}{|c|}{QNMs}&\multicolumn{2}{c|}{Relative deviations}\\
		\hline
		$L$&$\omega (SD)$&$\omega (WKB)$&$\left | Re \omega  \right |$&$\left | Im \omega  \right |$\\
		\hline
		1&0.2126877-0.0643820i&0.2082739-0.0665339i&2.119\%&3.234\%\\
		\hline
		3&0.4962712-0.0643820i&0.4943043-0.0647687i&0.398\%&0.597\%\\
		\hline
		5&0.7798548-0.0643820i&0.7785970-0.0645383i&0.162\%&0.242\%\\
		\hline
		7&1.0634383-0.0643820i&1.0625146-0.0644660i&0.087\%&0.130\%\\
		\hline
		10&1.4888137-0.0643820i&1.4881533-0.0644248i&0.044\%&0.066\%\\
		\hline
		15&2.1977726-0.0643820i&2.1973350-0.0644016i&0.020\%&0.030\%\\
		\hline
		20&2.9067315-0.0643820i&2.9063930-0.0643932i&0.012\%&0.017\%\\
		\hline
		50&7.1604848-0.0643820i&7.1603474-0.0643838i&0.002\%&0.003\%\\
		\hline
		
	\end{tabular}
\end{table}

\begin{table}[htbp]
	\centering
	\caption{The QNMs calculated by the shadow radius and sixth order WKB approximation method and the relative deviations, where $\omega _{0}=-\frac{2}{3}$, $\rho _{0}=0.1$, $\varepsilon ^{2}=0.001$.}
	\begin{tabular}{|c|c|c|c|c|}
		\hline
		&\multicolumn{2}{|c|}{QNMs}&\multicolumn{2}{c|}{Relative deviations}\\
		\hline
		$L$&$\omega (SD)$&$\omega (WKB)$&$\left | Re \omega  \right |$&$\left | Im \omega  \right |$\\
		\hline
		1&0.2170453-0.0661292i&0.2127676-0.0683394i&2.011\%&3.234\%\\
		\hline
		3&0.5064391-0.0661292i&0.5045253-0.0665271i&0.379\%&0.598\%\\
		\hline
		5&0.7958328-0.0661292i&0.7946085-0.0662901i&0.154\%&0.243\%\\
		\hline
		7&1.0852266-0.0661292i&1.0843273-0.0662157i&0.083\%&0.131\%\\
		\hline
		10&1.5193172-0.0661292i&1.5186743-0.0661733i&0.042\%&0.067\%\\
		\hline
		15&2.2428016-0.0661292i&2.2423658-0.0661494i&0.019\%&0.031\%\\
		\hline
		20&2.9662860-0.0661292i&2.9659565-0.0661408i&0.011\%&0.018\%\\
		\hline
		50&7.3071924-0.0661292i&7.3070586-0.0661311i&0.002\%&0.003\%\\
		\hline
		
	\end{tabular}
\end{table}

\begin{table}[htbp]
	\centering
	\caption{The QNMs calculated by the shadow radius and sixth order WKB approximation method and the relative deviations, where $\omega _{0}=-\frac{2}{3}$, $\rho _{0}=0.01$, $\varepsilon ^{2}=0.1$.}
	\begin{tabular}{|c|c|c|c|c|}
		\hline
		&\multicolumn{2}{|c|}{QNMs}&\multicolumn{2}{c|}{Relative deviations}\\
		\hline
		$L$&$\omega (SD)$&$\omega (WKB)$&$\left | Re \omega  \right |$&$\left | Im \omega  \right |$\\
		\hline
		1&0.2395636-0.0750453i&0.2405630-0.0762912i&0.415\%&1.633\%\\
		\hline
		3&0.5589818-0.0750453i&0.5593759-0.0752690i&0.070\%&0.297\%\\
		\hline
		5&0.8783999-0.0750453i&0.8786471-0.0751360i&0.028\%&0.121\%\\
		\hline
		7&1.1978180-0.0750453i&1.1979984-0.0750941i&0.015\%&0.065\%\\
		\hline
		10&1.6769453-0.0750453i&1.6770737-0.0750702i&0.008\%&0.033\%\\
		\hline
		15&2.4754906-0.0750453i&2.4755775-0.0750567i&0.004\%&0.015\%\\
		\hline
		20&3.2740360-0.0750453i&3.2741017-0.0750518i&0.002\%&0.009\%\\
		\hline
		50&8.0653082-0.0750453i&8.0653348-0.0750463i&0.001\%&0.001\%\\
		\hline
		
	\end{tabular}
\end{table}

\begin{table}[htbp]
	\centering
	\caption{The QNMs calculated by the shadow radius and sixth order WKB approximation method and the relative deviations, where $\omega _{0}=-\frac{2}{3}$, $\rho _{0}=0.001$, $\varepsilon ^{2}=0.1$.}
	\begin{tabular}{|c|c|c|c|c|}
		\hline
		&\multicolumn{2}{|c|}{QNMs}&\multicolumn{2}{c|}{Relative deviations}\\
		\hline
		$L$&$\omega (SD)$&$\omega (WKB)$&$\left | Re \omega  \right |$&$\left | Im \omega  \right |$\\
		\hline
		1&0.2457899-0.0776535i&0.2475929-0.0787969i&0.728\%&1.451\%\\
		\hline
		3&0.5735097-0.0776535i&0.5742592-0.0778572i&0.131\%&0.262\%\\
		\hline
		5&0.9012295-0.0776535i&0.9017036-0.0777361i&0.053\%&0.106\%\\
		\hline
		7&1.2289493-0.0776535i&1.2292964-0.0776979i&0.028\%&0.057\%\\
		\hline
		10&1.7205290-0.0776535i&1.7207766-0.0776762i&0.014\%&0.029\%\\
		\hline
		15&2.5398286-0.0776535i&2.5399962-0.0776639i&0.007\%&0.013\%\\
		\hline
		20&3.3591281-0.0776535i&3.3592548-0.0776595i&0.004\%&0.008\%\\
		\hline
		50&8.2749254-0.0776535i&8.2749768-0.0776545i&0.001\%&0.001\%\\
		\hline
		
	\end{tabular}
\end{table}

\begin{table}[htbp]
	\centering
	\caption{The QNMs calculated by the shadow radius and sixth order WKB approximation method and the relative deviations, where $\omega _{0}=-\frac{2}{3}$, $\rho _{0}=0.0001$, $\varepsilon ^{2}=0.1$.}
	\begin{tabular}{|c|c|c|c|c|}
		\hline
		&\multicolumn{2}{|c|}{QNMs}&\multicolumn{2}{c|}{Relative deviations}\\
		\hline
		$L$&$\omega (SD)$&$\omega (WKB)$&$\left | Re \omega  \right |$&$\left | Im \omega  \right |$\\
		\hline
		1&0.2464067-0.0779134i&0.2482913-0.0790460i&0.759\%&1.433\%\\
		\hline
		3&0.5749489-0.0779134i&0.5757346-0.0781115i&0.136\%&0.254\%\\
		\hline
		5&0.9034912-0.0779134i&0.9039884-0.0779951i&0.055\%&0.105\%\\
		\hline
		7&1.2320334-0.0779134i&1.2323974-0.0779574i&0.030\%&0.056\%\\
		\hline
		10&1.7248468-0.0779134i&1.7251065-0.0779359i&0.015\%&0.029\%\\
		\hline
		15&2.5462024-0.0779134i&2.5463782-0.0779237i&0.007\%&0.013\%\\
		\hline
		20&3.3675579-0.0779134i&3.3676909-0.0779193i&0.004\%&0.008\%\\
		\hline
		50&8.2956915-0.0779134i&8.2957455-0.0779144i&0.001\%&0.001\%\\
		\hline
		
	\end{tabular}
\end{table}

\begin{table}[htbp]
	\centering
	\caption{The QNMs calculated by the shadow radius and sixth order WKB approximation method and the relative deviations, where $\omega _{0}=-\frac{1}{2}$, $\rho _{0}=0.1$, $\varepsilon ^{2}=0.1$.}
	\begin{tabular}{|c|c|c|c|c|}
		\hline
		&\multicolumn{2}{|c|}{QNMs}&\multicolumn{2}{c|}{Relative deviations}\\
		\hline
		$L$&$\omega (SD)$&$\omega (WKB)$&$\left | Re \omega  \right |$&$\left | Im \omega  \right |$\\
		\hline
		1&0.1949690-0.0564726i&0.1930429-0.0574665i&0.998\%&1.730\%\\
		\hline
		3&0.4549276-0.0564726i&0.4540802-0.0566494i&0.187\%&0.312\%\\
		\hline
		5&0.7148863-0.0564726i&0.7143448-0.0565441i&0.076\%&0.126\%\\
		\hline
		7&0.9748449-0.0564726i&0.9744474-0.0565110i&0.041\%&0.068\%\\
		\hline
		10&1.3647829-0.0564726i&1.3644987-0.0564922i&0.021\%&0.035\%\\
		\hline
		15&2.0146795-0.0564726i&2.0144869-0.0564816i&0.010\%&0.016\%\\
		\hline
		20&2.6645761-0.0564726i&2.6644305-0.0564777i&0.005\%&0.009\%\\
		\hline
		50&6.5639558-0.0564726i&6.5638967-0.0564734i&0.001\%&0.001\%\\
		\hline
		
	\end{tabular}
\end{table}

\begin{table}[htbp]
	\centering
	\caption{The QNMs calculated by the shadow radius and sixth order WKB approximation method and the relative deviations, where $\omega _{0}=-\frac{1}{2}$, $\rho _{0}=0.1$, $\varepsilon ^{2}=0.01$.}
	\begin{tabular}{|c|c|c|c|c|}
		\hline
		&\multicolumn{2}{|c|}{QNMs}&\multicolumn{2}{c|}{Relative deviations}\\
		\hline
		$L$&$\omega (SD)$&$\omega (WKB)$&$\left | Re \omega  \right |$&$\left | Im \omega  \right |$\\
		\hline
		1&0.2330834-0.0717556i&0.2324959-0.0731014i&0.253\%&1.841\%\\
		\hline
		3&0.5438613-0.0717556i&0.5435746-0.0719967i&0.053\%&0.335\%\\
		\hline
		5&0.8546392-0.0717556i&0.8544532-0.0718533i&0.022\%&0.136\%\\
		\hline
		7&1.1654171-0.0717556i&1.1652799-0.0718082i&0.012\%&0.073\%\\
		\hline
		10&1.6315839-0.0717556i&1.6314856-0.0717824i&0.006\%&0.037\%\\
		\hline
		15&2.4085286-0.0717556i&2.4084619-0.0717679i&0.003\%&0.017\%\\
		\hline
		20&3.1854733-0.0717556i&3.1854228-0.0717627i&0.002\%&0.010\%\\
		\hline
		50&7.8471416-0.0717556i&7.8471211-0.0717568i&0.000\%&0.002\%\\
		\hline
		
	\end{tabular}
\end{table}

\begin{table}[htbp]
	\centering
	\caption{The QNMs calculated by the shadow radius and sixth order WKB approximation method and the relative deviations, where $\omega _{0}=-\frac{1}{2}$, $\rho _{0}=0.1$, $\varepsilon ^{2}=0.001$.}
	\begin{tabular}{|c|c|c|c|c|}
		\hline
		&\multicolumn{2}{|c|}{QNMs}&\multicolumn{2}{c|}{Relative deviations}\\
		\hline
		$L$&$\omega (SD)$&$\omega (WKB)$&$\left | Re \omega  \right |$&$\left | Im \omega  \right |$\\
		\hline
		1&0.2369897-0.0733726i&0.2365640-0.0747570i&0.180\%&1.852\%\\
		\hline
		3&0.5529760-0.0733726i&0.5527574-0.0736208i&0.040\%&0.337\%\\
		\hline
		5&0.8689623-0.0733726i&0.8688196-0.0734731i&0.016\%&0.137\%\\
		\hline
		7&1.1849486-0.0733726i&1.1848430-0.0734267i&0.009\%&0.074\%\\
		\hline
		10&1.6589280-0.0733726i&1.6588523-0.0734002i&0.005\%&0.038\%\\
		\hline
		15&2.4488937-0.0733726i&2.4488423-0.0733853i&0.002\%&0.017\%\\
		\hline
		20&3.2388594-0.0733726i&3.2388205-0.0733798i&0.001\%&0.010\%\\
		\hline
		50&7.9786537-0.0733726i&7.9786379-0.0733738i&0.000\%&0.002\%\\
		\hline
		
	\end{tabular}
\end{table}

\begin{table}[htbp]
	\centering
	\caption{The QNMs calculated by the shadow radius and sixth order WKB approximation method and the relative deviations, where $\omega _{0}=-\frac{1}{2}$, $\rho _{0}=0.01$, $\varepsilon ^{2}=0.1$.}
	\begin{tabular}{|c|c|c|c|c|}
		\hline
		&\multicolumn{2}{|c|}{QNMs}&\multicolumn{2}{c|}{Relative deviations}\\
		\hline
		$L$&$\omega (SD)$&$\omega (WKB)$&$\left | Re \omega  \right |$&$\left | Im \omega  \right |$\\
		\hline
		1&0.2414623-0.0757710i&0.2429172-0.0768937i&0.599\%&1.460\%\\
		\hline
		3&0.5634120-0.0757710i&0.5640122-0.0759709i&0.106\%&0.263\%\\
		\hline
		5&0.8853617-0.0757710i&0.8857409-0.0758520i&0.043\%&0.107\%\\
		\hline
		7&1.2073114-0.0757710i&1.2075888-0.0758146i&0.023\%&0.058\%\\
		\hline
		10&1.6902359-0.0757710i&1.6904338-0.0757932i&0.012\%&0.029\%\\
		\hline
		15&2.4951102-0.0757710i&2.4952441-0.0757812i&0.005\%&0.013\%\\
		\hline
		20&3.2999844-0.0757710i&3.3000857-0.0757768i&0.003\%&0.008\%\\
		\hline
		50&8.1292300-0.0757710i&8.1292710-0.0757719i&0.001\%&0.001\%\\
		\hline
		
	\end{tabular}
\end{table}

\begin{table}[H]
	\centering
	\caption{The QNMs calculated by the shadow radius and sixth order WKB approximation method and the relative deviations, where $\omega _{0}=-\frac{1}{2}$, $\rho _{0}=0.001$, $\varepsilon ^{2}=0.1$.}
	\begin{tabular}{|c|c|c|c|c|}
		\hline
		&\multicolumn{2}{|c|}{QNMs}&\multicolumn{2}{c|}{Relative deviations}\\
		\hline
		$L$&$\omega (SD)$&$\omega (WKB)$&$\left | Re \omega  \right |$&$\left | Im \omega  \right |$\\
		\hline
		1&0.2459750-0.0777249i&0.2478244-0.0788554i&0.746\%&1.434\%\\
		\hline
		3&0.5739417-0.0777249i&0.5747122-0.0779261i&0.134\%&0.258\%\\
		\hline
		5&0.9019084-0.0777249i&0.9023961-0.0778064i&0.054\%&0.105\%\\
		\hline
		7&1.2298751-0.0777249i&1.2302321-0.0777688i&0.029\%&0.056\%\\
		\hline
		10&1.7218252-0.0777249i&1.7220799-0.0777473i&0.015\%&0.029\%\\
		\hline
		15&2.5417419-0.0777249i&2.5419143-0.0777352i&0.007\%&0.013\%\\
		\hline
		20&3.3616587-0.0777249i&3.3617890-0.0777308i&0.004\%&0.008\%\\
		\hline
		50&8.2811591-0.0777249i&8.2812120-0.0777259i&0.001\%&0.001\%\\
		\hline
		
	\end{tabular}
\end{table}

\begin{table}[H]
	\centering
	\caption{The QNMs calculated by the shadow radius and sixth order WKB approximation method and the relative deviations, where $\omega _{0}=-\frac{1}{2}$, $\rho _{0}=0.0001$, $\varepsilon ^{2}=0.1$.}
	\begin{tabular}{|c|c|c|c|c|}
		\hline
		&\multicolumn{2}{|c|}{QNMs}&\multicolumn{2}{c|}{Relative deviations}\\
		\hline
		$L$&$\omega (SD)$&$\omega (WKB)$&$\left | Re \omega  \right |$&$\left | Im \omega  \right |$\\
		\hline
		1&0.2464251-0.0779205i&0.2483145-0.0790518i&0.761\%&1.431\%\\
		\hline
		3&0.5749920-0.0779205i&0.5757798-0.0781219i&0.137\%&0.258\%\\
		\hline
		5&0.9035589-0.0779205i&0.9040575-0.0780021i&0.055\%&0.105\%\\
		\hline
		7&1.2321257-0.0779205i&1.2324908-0.0779644i&0.030\%&0.056\%\\
		\hline
		10&1.7249760-0.0779205i&1.7252365-0.0779430i&0.015\%&0.029\%\\
		\hline
		15&2.5463932-0.0779205i&2.5465695-0.0779308i&0.007\%&0.013\%\\
		\hline
		20&3.3678104-0.0779205i&3.3679437-0.0779264i&0.004\%&0.008\%\\
		\hline
		50&8.2963134-0.0779205i&8.2963675-0.0779215i&0.001\%&0.001\%\\
		\hline
	\end{tabular}
\end{table}

\section{Conclusions}

In this paper, we investigated the photon sphere, shadow radius and QNMs of a black hole with a deficit solid angle and quintessence-like matter. The QNMs was derived by the sixth order WKB approximation approach and shadow radius. The values of the imaginary and real parts of the QNMs increase with the decrease of values of the density of quintessence-like matter and deficit solid angle. For the large values of the multiple number $L$, the values of the QNMs obtained by the two methods are consistent, which shows that the correspondence between the QNMs in the eikonal limit and shadow. When $L$=1, the real and imaginary parts derived by two methods is relatively large. However, when $L$$\ge 3$, the gap between them is significantly reduced. This investigation reveals the potential relationship between the black hole shadows and gravitational waves.

\vspace*{2.0ex}
\noindent \textbf{Acknowledgments}

\noindent This work is supported by Sichuan high-level talent support plan and the FXHU (Z201021).


\begin{thebibliography}{99}
	
\small
	
\bibitem{BPA}
B. P. Abbott et al, (LIGO Scientific Collaboration and Virgo Collaboration), \emph{Observation of gravitational waves from a binary black hole merger,} \emph{Phys. Rev. Lett.} \textbf{116} (2016) 061102.
	
B. P. Abbott et al, (LIGO Scientific Collaboration and Virgo Collaboration,) \emph{Tests of general relativity with GW150914}, \emph{Phys. Rev. Lett.} \textbf{116} (2016) 221101.
	
B. P. Abbott et al, (LIGO Scientific Collaboration and Virgo Collaboration,) \emph{GW151226: Observation of gravitational waves from a 22-solar-mass binary black hole coalescence}, \emph{Phys. Rev. Lett.} \textbf{116} (2016) 241103.
	
\bibitem{JMM}
J. M. Maldacena, \emph{The large N limit of superconformal field theories and supergravity,} \emph{Adv. Theor. Math. Phys.} \textbf{2} (1998) 231 [\emph{Int. J. Theor. Phys.} \textbf{38} (1999) 1113].
	
\bibitem{HH1}
D. T. Son and A. O. Starinets, \emph{Minkowski-space correlators in AdS/CFT correspondence: recipe and applications,} \emph{JHEP} \textbf{0209} (2002) 042.
	
\bibitem{HH2}
G. T. Horowitz and V. Hubeny, \emph{Quasinormal modes of AdS black holes and the approach to thermal equilibrium,} \emph{Phys. Rev.} \textbf{D 62} (2000) 024027.
	
\bibitem{KSS}
J. Morgan, A. S. Miranda and V. T. Zamchin, \emph{Electromagnetic quasinormal modes of rotating black strings and the AdS/CFT correspondence,} \emph{JHEP} \textbf{1303} (2013) 169.
	
\bibitem{MM}
P. Kovtun, D. T. Son and A. O. Starinets, \emph{Viscosity in strongly interacting quantum field theories from black hole physics,} \emph{Phys. Rev. Lett.} \textbf{94} (2005) 111601.
	
\bibitem{BBM}
H. J. Blome and B. Mashhoon, \emph{Quasi-normal oscillations of a Schwarzschild black hole,} \emph{Phys. Lett.} \textbf{A 100} (1984) 231
	
\bibitem{KDBF}
K. D. Kokkotas and B. F. Schutz, \emph{Black hole normal modes: A WKB approach. III. The Reissner-Nordstr$\ddot{o}$m black hole,} \emph{Phys. Rev.} \textbf{D 37} (1988) 3378.
	
\bibitem{ESSI}
E. Seidel and S. Iyer, \emph{Black hole normal modes: A WKB approach. IV. Kerr black holes,} \emph{Phys. Rev.} \textbf{D 41} (1990) 374.
	
\bibitem{HPN1}
H. P. Nollert, \emph{Quasinormal modes of Schwarzschild black holes: the determination of quasinormal frequencies with very large imaginary parts,} \emph{Phys. Rev.} \textbf{D 47} (1993) 5253.
	
\bibitem{HPN2}
S. B. Chen and J. L. Jing, \emph{Quasinormal modes of a black hole in the deformed Hořava-Lifshitz gravity,} \emph{Phys. Lett.} \textbf{B 687} (2010) 124.

\bibitem{CCDN1}
H. T. Cho, A. S. Cornell, J. Doukas, T. R. Huang and W. Naylor, \emph{A new approach to black hole quasinormal modes: A review of the asymptotic iteration method,} \emph{Adv. Math. Phys.} \textbf{2012} (2012) 281705.
	
\bibitem{CCDN2}
H. T. Cho, A. S. Cornell, J. Doukas and W. Naylor, \emph{Black hole quasinormal modes using the asymptotic iteration method,} \emph{Class. Quant. Grav.} \textbf{27} (2010) 155004.
	
\bibitem{LLQ}
K. Lin and W. L. Qian, \emph{A matrix method for quasinormal modes: Schwarzschild black holes in asymptotically flat and (anti-) de Sitter spacetimes,} \emph{Class. Quant. Grav.} \textbf{34} (2017) 095004.
	
\bibitem{BKK1}
J. L. Jing and Q. Y. Pan, \emph{Quasinormal modes and second order thermodynamic phase transition for Reissner-Nordstr$\ddot{o}$m black hole,} \emph{Phys. Lett.} \textbf{B 660} (2008) 13.
	
\bibitem{BKK2}
H. T. Cho, \emph{Dirac quasi-normal modes in Schwarzschild black hole spacetimes,} \emph{Phys. Rev.} \textbf{D 68} (2003) 024003.
	
\bibitem{BKK3}
J. S. F. Chan and R. B. Mann, \emph{Scalar wave falloff in asymptotically anti-de Sitter backgrounds,} \emph{Phys. Rev.} \textbf{D 55} (1997) 7546.
	
\bibitem{BKK4}
V. Cardoso and J. P. S. Lemos, \emph{Scalar, electromagnetic and weyl perturbations of BTZ black holes: quasinormal modes,} \emph{Phys. Rev.} \textbf{D 64} (2001) 124015.
	
\bibitem{BKK5}
J. L. Jing, \emph{Neutrino quasinormal modes of the Reissner-Nordstr$\ddot{o}$m black hole,} \emph{JHEP} \textbf{0512} (2005) 005.
	
\bibitem{BKK6}
F. W. Shu and Y. G. Shen, \emph{Quasinormal modes of Rarita-Schwinger field in Reissner-Nordstr$\ddot{o}$m black hole,} \emph{Phys. Lett.} \textbf{B 614} (2005) 195.
	
\bibitem{BKK7}
B. Wang, C. Y. Lin and E. Abdalla, \emph{Quasinormal modes of Reissner-Nordstr$\ddot{o}$m anti-de Sitter black holes,} \emph{Phys. Lett.} \textbf{B 481} (2000) 79.

\bibitem{RVA}
R. V. Konoplya and A. {Zhidenko},
\emph{Quasinormal modes of black holes: From astrophysics to string theory}
\emph{Rev. Mod. Phys.} \textbf{83} (2011) 793.
	
\bibitem{EW}
E. W. Leaver, \emph{An analytic representation for the quasi-normal modes of Kerr black holes,} \emph{Proc. Roy. Soc. Lond.} \textbf{A 402} (1985) 285

\bibitem{EW1}
E. W. Leaver, \emph{Spectral decomposition of the perturbation response of the Schwarzschild geometry,} \emph{Phys. Rev.} \textbf{D 34} (1986) 384

\bibitem{EW2}
E. W. Leaver, \emph{Quasinormal modes of Reissner-Nordström black holes,} \emph{Phys. Rev.} \textbf{D 41} (1990) 2986

\bibitem{NFPO}
N. Fröman, P. O. Fröman, N. Andersson and A. Hökback, \emph{Black hole normal modes: Phase-integral treatment,} \emph{Phys. Rev.} \textbf{D 45} (1992) 2609

\bibitem{NASL}
N. Andersson and S. Linnæus, \emph{Quasinormal modes of a Schwarzschild black hole: Improved phase-integral treatment,} \emph{Phys. Rev.} \textbf{D 46} (1992) 4179

\bibitem{DVAA}
D. V. Gal'tsov and A. A. Matiukhin, \emph{Matrix WKB method for black hole normal modes and quasibound states,} \emph{Class. Quant. Grav.} \textbf{9} (1992) 2039

\bibitem{BMNP}
B. Majumdar and N. Panchapakesan, \emph{Schwarzschild black hole normal modes using the Hill determinant,} \emph{Phys. Rev.} \textbf{D 40} (1989) 2568

\bibitem{SW}
B. F. Schutz and C. M. Will, \emph{Black hole normal modes: A semianalytic approach,} \emph{Astrophys.} \textbf{J. 291} (1985) L33.

\bibitem{IW}
S. Iyer and C. M. Will, \emph{Black hole normal modes: A WKB approach. I. Foundations and application of a higher order WKB analysis of potential-barrier scattering,} \emph{Phys. Rev.} \textbf{D 35} (1987) 3621.

\bibitem{IW1}
R. A. Konoplya, A. Zhidenko and A. F. Zinhailo, \emph{Higher order WKB formula for quasinormal modes and greybody factors: recipes for quick and accurate calculations,} \emph{Class. Quant. Grav.} \textbf{36} (2019) 155002.

R. Konoplya, \emph{Quasinormal behavior of the D-dimensional Schwarzshild black hole and higher order WKB approach,} \emph{Phys. Rev.} \textbf{D 68} (2003) 024018.

R. A. Konoplya, \emph{On quasinormal modes of small Schwarzschild-anti-de-Sitter black hole,} \emph{Phys. Rev.} \textbf{D 66} (2002) 044009.

\bibitem{IW2}
J. Matyjasek and M. Opala, \emph{Quasinormal modes of black holes. The improved semianalytic approach,} \emph{Phys. Rev.} \textbf{D 96} (2017) 024011.

\bibitem{CMBWZ}
V. Cardoso, A. S. Miranda, E. Berti, H. Witek and V. T. Zanchin, \emph{Geodesic stability, Lyapunov exponents and quasinormal modes,} \emph{Phys. Rev.} \textbf{D 79} (2009) 064016.

\bibitem{RAZS}
R. A. Konoplya and Z. Stuchlík, \emph{Are eikonal quasinormal modes linked to the unstable circular null geodesics?} \emph{Phys. Rev.} \textbf{B 771} (2017) 597.

\bibitem{RAAF}
R. A. Konoplya and A. F. Zinhailo, \emph{Quasinormal modes, stability and shadows of a black hole in the $4$D Einstein-Gauss-Bonnet gravity,} \emph{Phys. J.} \textbf{C 80} (2020) 1049.

\bibitem{EHT}
The Event Horizon Telescope Collaboration et al, \emph{First M87 event horizon telescope results. I. The shadow of the supermassive black hole,} \emph{Astrophys. J. Lett.} \textbf{875} (2019) L1.

The Event Horizon Telescope Collaboration et al, \emph{First M87 event horizon telescope results. IV. Imaging the central supermassive black hole,} \emph{Astrophys. J. Lett.} \textbf{875} (2019) L4.

The Event Horizon Telescope Collaboration et al, \emph{First M87 event horizon telescope results. V. Physical origin of the asymmetric ring,} \emph{Astrophys. J. Lett.} \textbf{875} (2019) L5.

The Event Horizon Telescope Collaboration et al, \emph{First M87 event horizon telescope results. VI. The shadow and mass of the central black hole,} \emph{Astrophys. J. Lett.} \textbf{875} (2019) L6.

\bibitem{JAAM1}
K. Jusufi, \emph{Quasinormal modes of black holes surrounded by dark matter and their connection with the shadow radius,} \emph{Phys. Rev.} \textbf{D 101} (2020) 084055.

\bibitem{JAAM2}
K. Jusufi, \emph{Connection between the shadow radius and quasinormal modes in rotating spacetimes,} \emph{Phys. Rev.} \textbf{D 101} (2020) 124063.

\bibitem{JAAM3}
B. Cuadros-Melgar, R. D. B. Fontana and J. de Oliveira, \emph{Analytical correspondence between shadow radius and black hole quasinormal frequencies,} \emph{Phys. Lett.} \textbf{B 811C} (2020) 135966.

\bibitem{JAAM4}
K. Jusufi, M. Amir, M. S. Ali and S. D. Maharaj, \emph{Quasinormal modes, shadow and greybody factors of 5D electrically charged Bardeen black holes,} \emph{Phys. Rev.} \textbf{D 102} (2020) 064020.

\bibitem{GM}
Y. Guo and Y. G. Miao, \emph{Null geodesics, quasinormal modes and the correspondence with shadows in high-dimensional Einstein-Yang-Mills spacetimes,} \emph{Phys. Rev.} \textbf{D 102} (2020) 084057.

\bibitem{MBAV}
M. Barriola and A. Vilenkin, \emph{Gravitational field of a global monopole,} \emph{Phys. Rev. Lett.} \textbf{63} (1989) 341.

\bibitem{BL1}
V. V. Kiselev, 
\emph{Quintessence and black holes,}
\emph{Class. Quant. Grav.} \textbf{20} (2003) 1187.

\bibitem{BL4}
A. D. Chernin, D. I. Santiago and A. S. Silbergleit, 
\emph{Interplay between gravity and quintessence: A Set of new GR solutions,}
\emph{Phys. Lett.} \textbf{A 294} (2002) 79.

\bibitem{BTZB}
B. {Toshmatov}, Z. {Stuchlk} and B. {Ahmedov},
\emph{Rotating black hole solutions with quintessential energy,} \emph{Eur. Phys. J. Plus} \textbf{132} (2017) 98.

\bibitem{BL3}
S. B. Chen, B. Wang and R. K. Su, 
\emph{Hawking radiation in a $d$-dimensional static spherically-symmetric black Hole surrounded by quintessence,}
\emph{Phys. Rev.} \textbf{D 77} (2008) 124011.

\bibitem{TD2}
M.A. A\"inou and M.E. Rodrigues, 
\emph{Thermodynamical, geometrical and Poincaré methods for charged black holes in presence of quintessence,}
\emph{JHEP} \textbf{09} (2013) 146.

\bibitem{OPLA}
O. {Pedraza}, L. A. {Lpez}, R. {Arceo} and I. C. {Munguia},
\emph{Geodesics of Hayward black hole surrounded by quintessence,} \emph{Gen. Rel. Grav.} \textbf{53} (2021) 24.

\bibitem{OPLA2}
O. {Pedraza}, L. A. {Lpez}, R. {Arceo} and I. C. {Munguia},
\emph{Quasinormal modes of the Hayward black hole surrounded by quintessence: Scalar, electromagnetic and gravitational perturbations,} \emph{Mod. Phys. Lett.} \textbf{A 37} (2022) 2250057.

\bibitem{BL2}
X. Ping, 
\emph{Quasinormal modes of a black hole with quintessence-like matter and a deficit solid angle,}
\emph{Astrophys. Space Sci.} \textbf{321} (2009) 47.

\bibitem{HAC}
H. {Chakrabarty}, A. {Abdujabbarov} and C. {Bambi},
\emph{Scalar perturbations and quasi-normal modes of a nonlinear magnetic-charged black hole surrounded by quintessence,} \emph{Eur. Phys. J.} \textbf{C 79} (2019) 179.

\bibitem{CXYQ}
C. X. {Sun}, Y. Q. {Liu}, W.L. {Qian} and R. H. {Yue},
\emph{Shadows of magnetically charged rotating black holes surrounded by quintessence,} arXiv:2201.01890.



	
	
	
\end{thebibliography}
\end{document}